\documentclass[aps,prl,twocolumn]{revtex4}
\usepackage{amsmath}
\usepackage{graphicx}
\usepackage{multirow}
\usepackage{array}
\usepackage{inputenc}
\begin{document}
 \author{X. Fabr\`eges$^1$, S. Petit$^1$, I. Mirebeau$^1$, S. Pailh\`es$^1$,
  L. Pinsard$^2$, A. Forget$^3$, M. T.  Fernandez-Diaz$^4$ and F. Porcher$^1$}
  \date{01/03/2009}
  \affiliation{$^1$ %Institut Rayonnement Mati\`ere de Saclay,
  Laboratoire L\'eon Brillouin, CEA-CNRS, CE-Saclay,
  91191 Gif-sur-Yvette,France}
  \affiliation{$^2$Laboratoire de Physico-Chimie de l'Etat Solide,ICMMO, Universit\'e Paris-Sud, 91405 Orsay, France}
  \affiliation{$^3$%Institut Rayonnement Mati\`ere de Saclay,
  Service de Physique de l'Etat Condens\'e, CEA-CNRS, CE-Saclay,
  91191 Gif-Sur-Yvette, France}
  \affiliation{$^4$Institut La\"ue Langevin, 6 rue Jules Horowitz, BP 156X, 38042
Grenoble France.}
  \title{Spin-lattice coupling, frustration and magnetic order in multiferroic RMnO$_3$}
  \begin{abstract}

    We have performed high resolution neutron diffraction and inelastic neutron scattering experiments in
    the frustrated multiferroic hexagonal compounds RMnO$_3$ (R=Ho, Yb, Sc, Y), which provide evidence
    of a strong magneto-elastic coupling in the the whole family.
    We can correlate the atomic positions, the type of magnetic structure and the nature
    of the spin waves whatever the R ion and temperature. The key parameter
    is the position of the Mn ions in the unit cell with respect to a critical threshold of $1/3$,
    which determines the sign of the coupling between Mn triangular planes.
\end{abstract}

\maketitle
%%%%%%%%%%%%%%%%%%%%%%%%%%%%%%%%
% Introduction generale du sujet

Multiferroics have aroused a great attention for the last years,
as the coupling between ferroelectric and magnetic orderings in these
materials may open the route to novel promising electronic devices.
Magnetic frustration combined with a striking magneto-elastic coupling
seems to be at the origin of their properties, a cocktail that has a
strong potential for novel physics\cite{cheong}.
These compounds are however rare, and far from being fully understood. Indeed,
ferroelectricity imposes a non centrosymmetric space group \cite{Fiebig05},
while magnetic frustration favors complex magnetic orders \cite{Greedan01,Moessner06}.

The RMnO$_3$ compounds, where R is a rare earth ion, follow these conditions.
In orthorhombic RMnO$_3$, (R=Eu, Gd, Tb, Dy), where the strong GdFeO$_3$-type distortion
lifts the orbital degeneracy, magnetic frustration arises from competing super-exchange
interactions \cite{Kimura03}, yielding incommensurate magnetic structures
\cite{Kenzelmann05}. This peculiar ordering suggested a novel coupling between
dielectric and magnetic collective modes \cite{Katsura07}. In hexagonal RMnO$_3$ 
with smaller R ionic radius, (R=Ho, Er, Yb, Lu, Y), magnetic frustration arises 
from the triangular geometry \cite{Jolicoeur89}, yielding  120$^\circ$ %\mathdegree
N\'eel orders for the Mn moments \cite{Munoz01}. In YMnO$_3$ and LuMnO$_3$, where Y and Lu are non magnetic,
an iso-structural transition was recently observed at the N\'eel temperature T$_N$ \cite{Lee08},
namely each ion "moves" inside the unit cell when Mn moments get ordered. This effect provided
evidence for a giant magneto-elastic coupling, likely connected with an increase
of the ferroelectric polarization \cite{Lee05}. Its origin remains unexplained so far.

To shed light on these materials, we carried out high resolution
neutron diffraction and inelastic neutron scattering in four RMnO$_3$ (R=Y, Sc,
Ho, Yb), with either non magnetic (Y, Sc) or magnetic (Ho, Yb) R ion, showing
(Ho, Sc) or not (Y,Yb) a spin reorientation at T$_{SR}$ with temperature
\cite{Munoz00,Brown06,Fabreges08}. We show that the iso-structural transition
is a systematic feature in the hexagonal series. In addition, we establish a
correlation between the atomic positions, the type of magnetic structure, and
the nature of the spin waves, whatever the compound and its magnetic structure.
We show that the key parameter is the position x of the Mn ions within the
triangular plane with respect to a critical threshold of $1/3$ which tunes 
the sign of  the interplane exchange interaction. We justify this result by simple 
energy arguments.
Thanks to the magneto-elastic coupling, the atomic motion helps releasing the
frustration by selecting a given magnetic structure, depending on the $x$ value.
This process recalls the spin-Peierls states stabilized
in several geometrically frustrated 2D or 3D compounds \cite{Becca02,Tchernyshyov02}.

%%%%%%%%%%%%%%%%%%%%%%%%%%%%%%%%
\begin{figure}
\begin{center}
        \includegraphics[width=8cm]{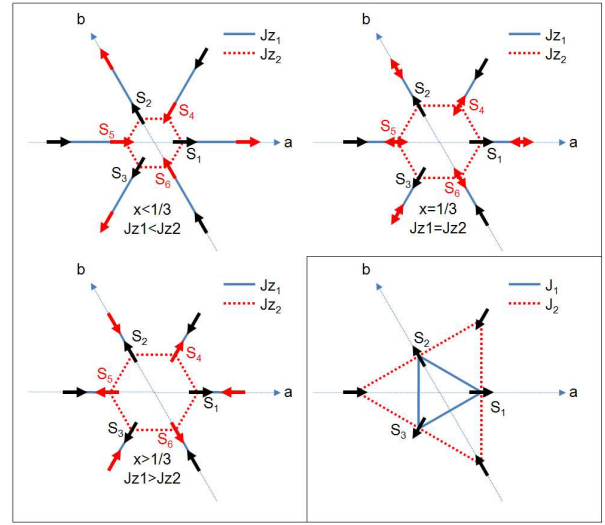}
  \caption{[Colors on line].
	Sketch of the hexagonal MnO planes with out-of-plane exchange
	paths $J_{z1}$ (between $\vec{S}_4$ and $\vec{S}_3$) and $J_{z2}$
	(between $\vec{S}_4$ and $\vec{S}_1$ as well as $\vec{S}_4$ and
	$\vec{S}_2$). Black (red) arrows depict the positions
	of the Mn spins within z=0 (z=1/2) Mn planes. $\vec{S}_1,\vec{S}_2$
	and $\vec{S}_3$ are located at $(x,0,0)$, $(0,x,0)$, $(-x,-x,0)$,
	while $\vec{S}_4,\vec{S}_5,\vec{S}_6$ at $(x,x,1/2)$, $(1-x,0,1/2)$
	and $(0,1-x,1/2)$. Double arrows are used when several spin
	orientations can be stabilized. Inset : sketch of the hexagonal MnO
	plane with in-plane exchange paths $J_{1}$ and $J_{2}$.} \label{fig1}
\end{center}
\end{figure}

%%%%%%%%%%%%%%%%%%%%%%%%%%%%%%%%
% Description des hexagonaux

Hexagonal RMnO$_3$ compounds consist of stacked Mn-O and R layers, the Mn ions
forming a nearly ideal two dimensional triangular lattice \cite{Katsufuji02}. 
They crystallize in the $P6_3cm$ space group, with two Mn-O planes per unit cell.
As shown on Figure \ref{fig1}, Mn coordinates depend on a unique
parameter $x$. For $x \ne 1/3$, two different exchange path $J_{1}$
and $J_{2}$ can be defined between Mn moments in a given Mn plane
(inset Fig. \ref{fig1}).
The triangular symmetry and hence the geometrical frustration within a
Mn plane is however preserved whatever the $x$ value. The Mn-Mn interactions
between adjacent Mn planes are due to super-super exchange paths, via the
apical oxygen ions of MnO$_5$ bipyramids. These interactions lead to a 3D
magnetic ordering below T$_N$. Again, as soon as $x \ne 1/3$, two different 
paths, and thus two different interactions, $J_{z1}$ and $J_{z2}$, can be 
distinguished (Fig.\ref{fig1}), while for $x=1/3$ all paths become equivalent.
The value $x =1/3$ is therefore a critical threshold which determines
the sign of the effective interplane exchange $J_{z1}-J_{z2}$ (the same sign 
as that of $x -1/3$). As shown below, this quantity determines in turn the 
stability of the magnetic phases and the nature of the spin waves.

\begin{figure}
\begin{center}
\includegraphics[width=8cm]{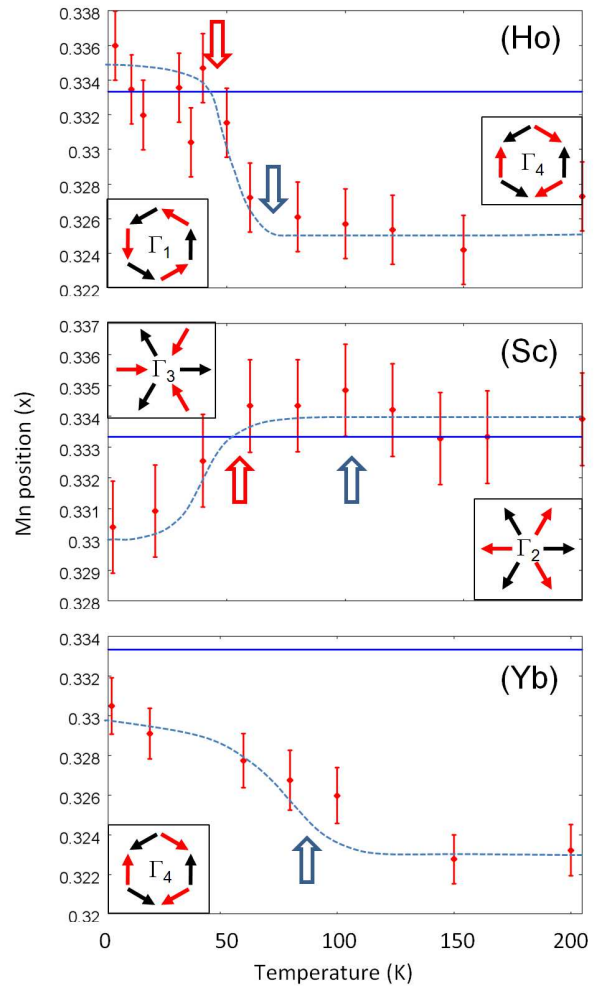}
\caption{[Color online]. Refined Mn positions in the unit cell
obtained from high-resolution neutron powder diffraction.
Blue and red arrows indicate respectivly the N\'eel
 and the spin reorientation temperatures.
Lines are guides to the eyes. Right insets : Mn magnetic configuration for
$T_{SR}<T<T_N$. Left insets : Mn magnetic configuration
$T<T_{SR}$).} \label{fig2}
\end{center}
\end{figure}

Whatever the $x$ value and the R magnetism, the Mn magnetic moments order
in 120$^\circ$ arrangements, with four different possible structures
(inset Fig. \ref{fig2}), labeled from the irreducible representations (IR)
$\Gamma_{i,~i=1-4}$ of the $P6_3cm$ space group with $\bf{k}$=0 propagation
vector \cite{Munoz00,Bertaut_gamma}. For $\Gamma_1$ and $\Gamma_4$, Mn moments are
perpendicular to $a$ and $b$ axes, and their arrangement in the $z=1/2$ plane
are either antiparallel ($\Gamma_1$) or parallel ($\Gamma_4$) with respect
to z=0. The same picture holds for $\Gamma_2$ and $\Gamma_3$ with spins
along $a$ and $b$ axes.

%%%%%%%%%%%%%%%%%%%%%%%%%%%%%%%%
% elastic

High-resolution neutron powder diffraction patterns were collected versus
temperature on the D2B and 3T2 instruments, at ILL and LLB-Orph\'ee reactors
respectively. Powder samples HoMnO$_3$, ScMnO$_3$ and YbMnO$_3$ were prepared
as described in \cite{Alonso_poudre} and characterized by x-ray diffraction.
Data were analyzed using the Fullprof and Basireps softwares
\cite{Juan_fullprof,Juan_basireps}, allowing us to determine the atomic positions
and the magnetic structures versus temperature precisely. The magnetic structures
and transition temperatures agree with previous determinations \cite{Munoz00,Munoz01}.
It is worth noting that HoMnO$_3$ and ScMnO$_3$ undergo a second magnetic
transition at T$_{SR}$ corresponding to the reorientation of the Mn spins.

Figure \ref{fig2} shows the temperature dependence of the Mn position $x$ for
the three samples. As a striking feature, in HoMnO$_3$, $x$ exhibits an
unprecedented large change (of about 3\%), which occurs at $T_N$,
providing evidence for an iso-structural transition concomitant with the
magnetic ordering. Smaller changes occur at $T_N$  in ScMnO$_3$ and YbMnO$_3$.
These results confirm those previously obtained \cite{Lee08} in Y and LuMnO$_3$,
and show that this transition is universal in the hexagonal RMnO$_3$ series. As
noticed in Ref.\onlinecite{Lee08}, the variations of $x$ versus $T$ depend of 
the rare-earth ion, namely $x$ increases below T$_N$ in Ho and YbMnO$_3$, whereas 
it decreases in ScMnO$_3$. Moreover, we also discern important changes 
at the reorientation transition. In HoMnO$_3$, $x$ increases with decreasing
temperature and crosses the 1/3 threshold exactly at the spin reorientation
temperature ($T_{SR}$). The reverse situation holds for ScMnO$_3$, where $x$
decreases upon cooling, becoming lower than 1/3 at T$_{SR}$. Finally, in
YbMnO$_3$, which shows no reorientation transition, $x$ increases at T$_N$
but remains lower than 1/3 in the whole temperature range.

%%%%%%%%%%%%%%%%%%%%%%%%%%%%%%%%%%%%%%
% 1ere discussion

All these features can be explained by considering the strategic
position of the Mn ions. We argue that $x$ is the key parameter that
controls the sign of $J_{z1} - J_{z2}$ and thus drives the magnetic
ordering. Namely for $x \le 1/3$, the exchange path along $J_{z1}$
is longer than along $J_{z2}$ and we expect $J_{z1} - J_{z2} \le 0$.
The reverse situation occurs for $x \ge 1/3$. A careful look at the
positions and magnetic structures summarized in table \ref{table1}
shows that $x \le 1/3$ is associated with $\Gamma_3$ and $\Gamma_4$
while $x \ge 1/3$ corresponds to $\Gamma_1$ and $\Gamma_2$.
This scheme also explains the occurrence of a spin reorientation
transition when $x$ crosses the 1/3 threshold. In this case,
$J_{z1} - J_{z2}$ changes sign, resulting in a change from $\Gamma_1$
towards $\Gamma_4$ (Ho) or from $\Gamma_2$ towards $\Gamma_3$ (Sc).
The crucial importance of the $J_{z1} - J_{z2}$ coupling can be
further justified by calculating the magnetic energy ${\cal E}$
(per unit cell) of the Mn moments in a mean field approximation.
${\cal E}$ is readely obtained by writting the Heisenberg Hamiltonian :
${\cal H} = {\cal H}_p + {\cal H}_z$,
with ${\cal H}_p= \sum{J~\vec{S}_{i} \vec{S}_{j}}$, and
${\cal H}_z = \sum{J_z~\vec{S}_{i} \vec{S}_{j}}$,
where the sums run over nearest neighbors and the subscripts p and
z refer to in plane and out of plane interactions respectively.
Because of the triangular arrangement, we have $\sum_{i=1,2,3}
\vec{S}_{i} = \sum_{i=4,5,6} \vec{S}_{i} = 0$, which in turn
implies : ${\cal E} = -\frac{3}{2} J S^2 + (J_{z1} - J_{z2})
(\vec{S}_3.\vec{S}_4)$. Depending on its sign, parallel or
antiparallel orientations of $\vec{S}_3$ and $\vec{S}_4$ are
expected, giving rise to the four magnetic structures described
above. Within this simple picture, $\Gamma_1$ and $\Gamma_2$ IR
minimize the energy for $J_{z1} - J_{z2} \ge 0$ while $\Gamma_3$
and $\Gamma_4$ are favored for $J_{z1} - J_{z2} \le 0$. This is
in exact agreement with the results summarized in Table 1.

\begin{table}
\begin{tabular}{c | c c | c c}
\multirow{2}{*}{R ion} & \multicolumn{2}{c |}{IR} & \multicolumn{2}{c}{Positions} \\
       & $T_N$ & 1.5K & $T_N$ & 1.5K \\
       \hline
       Yb & $\Gamma_4$ & $\Gamma_4$ & 0.3270(15)   & 0.3310(14)   \\
       Ho & $\Gamma_4$ & $\Gamma_1$ & 0.3261(21)   & 0.3359(19)   \\
       Sc & $\Gamma_2$ & $\Gamma_3$ & 0.3342(18)   & 0.3304(17)   \\
       Y  & $\Gamma_1$ & $\Gamma_1$ & 0.3330(17)   & 0.3423(13)
\end{tabular}
\caption{Mn position in RMnO$_3$ compounds correlated with their magnetic structures
defined by $\Gamma$ irreducible representations. YMnO$_3$ positions are taken from
Ref. \onlinecite{Lee08}} \label{table1}
\end{table}

%%%%%%%%%%%%%%%%%%%%%%%%%%%%%%%%%%%%%%%%%%%%%%%%%%%%%%%
% Inelastic
%%%%%%%%%%%%%%%%%%%%%%%%%%%%%%%%%%%%%%%%%%%%%%%%%%%%%%%

A straightforward way to confirm our explanation is to determine
the value of $J_{z1} - J_{z2}$ by an independent measurement. This
can be easely done by measuring the spin wave dispersion along the
c-axis. For this purpose we carried out inelastic neutron
scattering experiments on the cold triple-axis 4F spectrometer
installed at LLB-Orph\'ee, on large single crystals of YMnO$_3$,
YbMnO$_3$ and HoMnO$_3$ grown by the floating zone technique.

\begin{figure}
\begin{center}
\includegraphics[width=8cm]{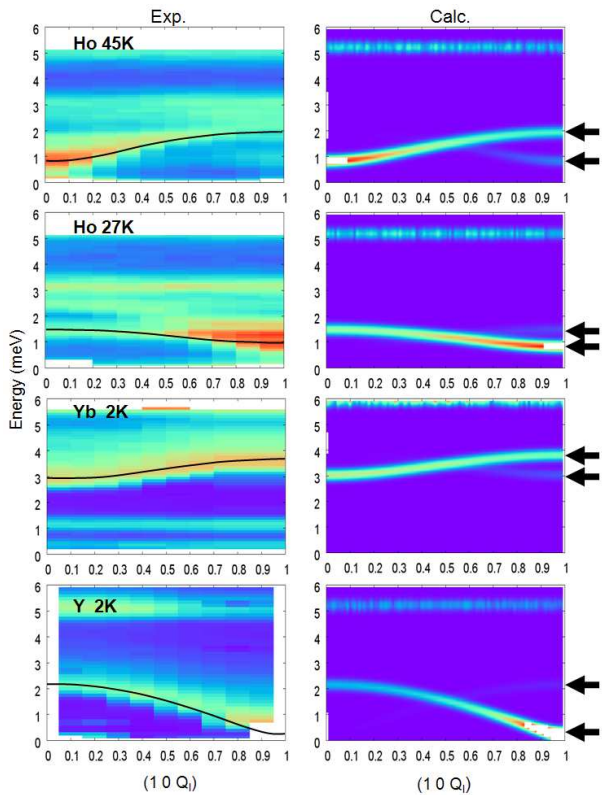}
\caption{(color online.) Maps of the dynamical
structure factor in Ho, Yb and YMnO$_3$. The neutron intensity is
plotted versus the energy transfer $\omega$ and wave vector $(1,0,Q_{\ell})$.
Left panel : experimental results from inelastic neutron scattering.
Black lines highlight the low energy spin wave modes.
Additional Q-independent lines in Yb and HoMnO$_3$ maps are due to
crystal field excitations.
Right panel : spin wave dispersion curves calculated using an Heisenberg
Hamiltonian. Black arrows indicate the two Goldstone modes (see text).}
\label{fig3}
\end{center}
\end{figure}

Figure \ref{fig3}(left) shows a color map of the dynamical structure
factor measured as a function of energy and wavevector in YMnO$_3$,
YbMnO$_3$ and HoMnO$_3$. These maps were obtained by collecting
energy scans taken at different $(1,0,Q_{\ell})$ wavevectors. The
measurements were performed at 2 K in the first two cases, and
at two temperatures just above and below the re-orientation
temperature T$_{SR}$ for HoMnO$_3$.
Different features can be seen from these experimental data,
including crystal field levels (Yb and Ho). A comprehensive 
investigation of of these features is however beyond the scope 
of this paper and we would like to focus on the low energy
spin wave excitations labelled with arrows. We notice that in YMnO$_3$
and YbMnO$_3$, this particular branch displays upwards (Yb) or
downwards (Y) dispersions, revealing opposite couplings along c.
Similarly in HoMnO$_3$, when crossing the reorientation T$_{SR}$,
the curvature changes from upwards to downwards, indicating a
reversal of the magnetic interaction along the c-axis.

To get a quantitative information about these couplings, we
performed a spin wave analysis of the Heisenberg Hamiltonian
${\cal H} = {\cal H}_p+{\cal H}_z$ defined above, taking into
account additional planar and uniaxial anisotropy terms \cite{sato,petit}.
Figure \ref{fig3}(right) shows the dynamical structure factors
calculated on the basis of this model as a function of energy
transfer $\omega$ and wavevector $(1,0,Q_{\ell})$. In this
approach, six spin wave modes are expected. Along c, four of
them are almost degenerate. They exhibit a large gap due
to the planar anisotropy, as well as a very weak dispersion.
On Figure \ref{fig3} (right), these modes correspond to
the flat branch sitting at 5 meV (Ho,Y) or 6 meV (Yb). The two
remaining modes correspond to the Goldstone modes of the magnetic
structure. In $q=0$ limit, they can be seen as global rotations
of the 120$^\circ$ pattern inside the basal planes. They couple
via the $J_z$ interactions, resulting in either in phase or out
of phase rotations. These two modes are pointed out by arrows in Figure
\ref{fig3} (right). Actually, one of them has a vanishing intensity
but can still be observed around $(1 0 1)$.
Assuming antiferromagnetic couplings $J_{z1}$ and $J_{z2}$, we
can determine from the data $J_{z1} - J_{z2} = 0.0050(5)$ meV in YMnO$_3$
and $J_{z1} - J_{z2} = -0.012(2)$ meV in YbMnO$_3$. For HoMnO$_3$,
the neutron data are well modelled with $J_{z1} - J_{z2} = -0.0038(5)$ meV 
at T=45K $\ge$ T$_{SR}$ and with $J_{z1} - J_{z2} = 0.0018(5)$ meV at 
T=27K $\le$ T$_{SR}$. We note that the corresponding signs of 
$J_{z1} - J_{z2}$ deduced from these measurements are fully consistent 
with the diffraction analysis.

%%%%%%%%%%%%%%%%%%%%%%%%%%%%%%%%%%%%%%%%%%%%
% discussion
%%%%%%%%%%%%%%%%%%%%%%%%%%%%%%%%%%%%%%%%%%%%

Our analysis emphasizes the double origin of the magnetic frustration. The Mn 
triangular planes are geometrically frustrated for antiferromagnetic interactions. 
In addition, adjacent Mn planes are coupled along the $c$ axis by self-competing
interactions. The Mn shift with respect to the 1/3 position does not suppress the 
rotational invariance in the Mn plane, but clearly lifts the interplane frustration, 
allowing 3D ordering. Depending on $x$, either ($\Gamma_1,\Gamma_2$) 
or ($\Gamma_3,\Gamma_4$) structures are stabilized. The remaining degree of freedom 
in the system is the global rotation of a 120$^\circ$ N\'eel order around the c-axis. 
Namely, to stabilize $\Gamma_1$ or $\Gamma_2$ ($\Gamma_3$ or $\Gamma_4$), the Mn 
magnetic moments must couple to the $(a,b)$ crystal axes.

On the one hand, residual anisotropic
interactions among Mn ions can in principle select a given orientation in a Mn
plane. As revealed by the exceptionally small uniaxial gap observed in YMnO$_3$
\cite{sato,petit}, these interactions are weak. They are not sufficient to break
the triangular symmetry, so that two-dimensional spin liquid fluctuations remain
\cite{sato}. On the other hand, one could argue that R-Mn interactions (when R
is magnetic) play a significant role. Indeed, the R moments on the $4b$ site order
at T$_N$ in the Mn molecular field, while their orientations are clearly coupled
to the Mn ones, through an energy term of anisotropic nature \cite{Fabreges08}.
Our spin wave measurements (Fig \ref{fig3}) show that the uniaxial anisotropy gap
in YbMnO$_3$ (3 meV) varies with temperature like the Yb moment, providing evidence
for such R-Mn coupling. Nevertheless, we shall not conclude that the Mn orientation
in the $(a,b)$ plane is completely determined by this interaction. For instance,
spin reorientation transitions of the Mn sublattice may occur (Sc, Ho) or not
(Y, Yb) whatever the R magnetism.

%%%%%%%%%%%%%%%%%%%%%%%%%%%%%%%%%%%%%%%%%%%%
% Resume
%%%%%%%%%%%%%%%%%%%%%%%%%%%%%%%%%%%%%%%%%%%%
In conclusion, our elastic and inelastic neutron scattering experiments clearly
show the crucial role of the Mn position in determining both the magnetic structure
and the spin waves modes. The onset of Mn magnetic orderings at T$_N$ or T$_{SR}$
correlates with the Mn position. %releasing the magnetic frustration at the same
%time.
The magnetic orders and spin excitations in the whole series result from a
subtle interplay of magneto-elastic coupling, frustrated intra- and inter-plane
Mn-Mn interactions and R-Mn interactions. The strong importance of inter-plane
interactions strongly suggests that the shift of the Mn position, which releases
the frustration along the $c$ axis, is a key ingredient at the origin of the
multiferroicity in the hexagonal RMnO$_3$ family.
We thank E. Suard for her help in the  D2B experiment and V. Simonet and 
F. Damay-Rowe for useful discussions.
%%%%%%%%%%%%%%%%%%%%%%%%%%%%%%%%%%%%%%%%%%%%

\end{document}